\documentclass[12pt]{article}
\usepackage{a4,epsf}

\newcommand \ket[1]{\left\vert\, {#1} \, \right>}
\newcommand \braket[2]{\hbox{$\left< {#1} \,\vrule\, {#2} \right>$}}
\newcommand{\bea}{\begin{eqnarray}}
\newcommand{\eea}{\end{eqnarray}}
\newcommand{\simgt}{\hbox{ \raise3pt\hbox to 0pt{$>$}\raise-3pt\hbox{$\sim$} }}
\newcommand{\simlt}{\hbox{ \raise3pt\hbox to 0pt{$<$}\raise-3pt\hbox{$\sim$} }}
\newcommand \vc[1]{{\bf {#1}}}

\begin{document}
\begin{titlepage}
\title{Top Quark Pair Production and Decay Near Threshold\\
in $e^+e^-$ Collisions\thanks{Presented 
at XXI School of Theoretical Physics, 
Ustro\'{n}, Poland, September, 1997.}
}
\author{Y.~Sumino\thanks{On 
leave of absence from Department of Physics, Tohoku University,
Sendai 980-77, Japan.}
\\ \\ Institut f\"ur Theoretische Teilchenphysik,
Universit\"at Karlsruhe,\\
D-76128 Karlsruhe, Germany
}
\date{}
\maketitle
\thispagestyle{empty}
\vspace{-3.8truein}
\begin{flushright}
{\bf TTP 97-45}\\
{\bf hep-ph/9711233}\\
{\bf November 1997}
\end{flushright}
\vspace{3.0truein}
\vspace{3cm}
\begin{abstract}
\noindent
{\small
We review the physics involved in the production and decay of top quarks
in $e^+e^- \to t\bar{t}$ near threshold,
with special emphasis on 
the recent theoretical study on the decay process of top
quarks in the threshold region.
The energy-angular distribution of $l^+$ in semileptonic top decays
is calculated including the full ${\cal O}(\alpha_s)$ corrections.
Various effects of the final-state interactions are elucidated.
A new observable is defined near threshold, which depends only on the
decay of {\it free} polarized top quarks, and thus it can be
calculated without bound-state
effects or the final-state interactions.
}
\end{abstract}
\vfil

\end{titlepage}
  
\section{Introduction}
A future $e^+e^-$ linear collider operating at energies
around the $t\bar{t}$ threshold will be one of the
ideal testing grounds for
unraveling the properties of the top quark. 
So far there have been a number
of studies of the cross section for top-quark pair production
near the $t\bar{t}$ threshold, both
theoretical and experimental \cite{nlc}--\cite{ps}, in which it has been
recognized
that this kinematical region is rich in physics and is also apt for 
extracting
various physical parameters efficiently.
The purpose of this paper is to review the physics involved
in the $t\bar{t}$ threshold region, with special emphasis on 
the recent theoretical
study \cite{ps} on the decay processes of top
quarks in this region.

After the introduction,
we discuss the physics concerning the production process of the
top quark in Section 2.
We assess the new results on the decay process of top quarks in the 
threshold region in Section 3.
A summary is given in Section 4.

\subsection{Top Quark Properties}
Let us first recall some basic properties of the top quark.
Its mass is now measured to around $\pm 5$~GeV
accuracy.
The recent reported values are
\bea
m_t = \left\{
\begin{array}{llllllc}
175.9 &\pm& 4.8 &\pm& 4.9&\mbox{GeV} & \mbox{(CDF \cite{cdf})}\\
173.3 &\pm& 5.6 &\pm& 6.2&\mbox{GeV} & \mbox{(D0 \cite{dzero})}
\end{array} .
\right.
\eea
Within the standard model,
the top quark decays almost 100\% to
$b$ quark and $W$.
The decay width of top quark $\Gamma_t$ is predictable 
as a function of $m_t$, and
already a fairly precise theoretical prediction
at the level of a few percent accuracy is available \cite{jk1}.
Here, we only note that $\Gamma_t \simeq 1.5$~GeV for the
above top quark mass range.
Another important property of the top quark 
is that it decays so quickly that
no top-hadrons will be formed.
Therefore all the spin information of the top quark
will be transferred to its decay daughters in its decay 
processes \cite{kuehn}, and
the energy-angular distributions of the decay products are
calculable as purely partonic processes.
In fact we may take full advantage of the spin information 
in studying the
top quark properties through its decay processes \cite{peskin}.

\subsection{$t\bar{t}$ System Near Threshold}
The $t\bar{t}$ production process near
threshold is considered as a candidate for the
first stage operation of next-generation linear $e^+e^-$
colliders (NLC), 
since the study of top quark threshold is quite promising and
also very interesting among the various subjects of NLC.
In analogy with charmonium or bottomonium production, one might
expect toponium resonance formations and accordingly
enhancement of the QCD interaction also in the $t\bar{t}$
threshold region.
There will appear, however, unique features to this system which make
it very different from the charmonium or bottomonium, as
we will see below.

Theoretically, quite stable 
predictions of cross sections are available near $t\bar{t}$ threshold
due to the following reasons.
First,
the large top quark mass allows us to probe the deep region of
the QCD potential, in the asymptotic regime 
where the strong coupling constant
$\alpha_s$ is small.
Secondly, the large width $\Gamma_t$ of the top quark acts as an infra-red
cut-off, which prevents hadronization effects affecting the cross section
\cite{fk}.
The toponium resonances decay
dominantly via electroweak
interaction\cite{kz,hkmn} so that 
their decay process can be calculated reliably.
Thirdly, the leading order QCD enhancement comes from the spacelike
region of the gluon momentum, hence the theoretical predictions are 
more stable in comparison to the predictions for timelike QCD 
processes.\footnote{
The threshold cross section is sensitive to $\alpha_s$ due to
an enhancement by the QCD interaction.
We may compare it with other physical
quantities which are also sensitive to $\alpha_s$, e.g.\ 
various semi-inclusive observables from jet 
physics, which involve timelike processes.
}

It is illuminating to consider
the time evolution of this system, 
a $t\bar{t}$ pair produced in $e^+e^-$
annihilation as they spread
apart from each other.
Since they are slow near the threshold, they cannot
escape
even relatively weak attractive force mediated by the exchange of
Coulomb gluons;
$t$ and $\bar{t}$ are bound to form Coulombic resonances
when they reach the distance of Bohr radius
$(\alpha_s m_t)^{-1} \sim 0.1$~GeV$^{-1}$.
At this stage, the coupling of top quark to gluon
is of the order of $\alpha_s(\mu \! = \! \alpha_sm_t) \sim 0.15$.
If they could continue to
spread apart even further to the distance
$\Lambda_{QCD}^{-1} \sim$ a few GeV$^{-1}$,
there would occur the hadronization
effects as the coupling becomes very strong,
since gluons with wave-length $\sim \Lambda_{QCD}^{-1}$ would be able to resolve
the color charge of each constituent.
For a realistic top quark, however, the $t\bar{t}$ pair will
decay at the distance $(m_t\Gamma_t)^{-1/2} \sim 0.1$~GeV$^{-1}$
into energetic $b$ and $\bar{b}$ jets and $W$'s
well before the hadronization
effects become important.
Thus, the toponium can be regarded as a Coulombic resonance
state (with reasonably weak coupling)
due to the large mass and the large width of the top quark.

\subsection{Theoretical Background}

\begin{figure}
\begin{center}
  \leavevmode
  \epsfxsize=10.cm
  \epsffile{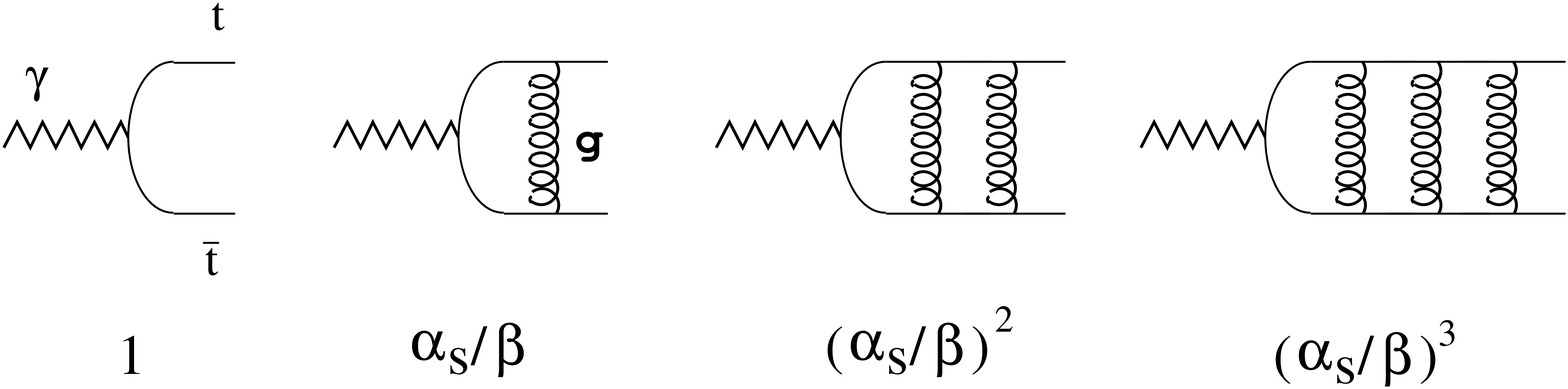}\vspace{5mm}\\  
  \caption[]{
        \label{ladder}The ladder diagrams for the process 
$\gamma^* \to t\bar{t}$.
The diagram where $n$ uncrossed gluons are exchanged has the behavior
$(\alpha_s/\beta)^n$ near threshold.
}
\end{center}
\end{figure}

Let us consider the amplitude for
$\gamma^* \to t\bar{t}$ at the c.m.\ energy slightly above the threshold.
It is well known that the ladder diagram for this
process where uncrossed
gluons are exchanged $n$ times between $t$ and $\bar{t}$
has the behavior $\sim (\alpha_s/\beta)^n$, see Fig.~\ref{ladder},
where $\beta = \sqrt{1-{4m_t^2}/{s}}$ is
the velocity of $t$ or $\bar{t}$ in the c.m.\ frame,
which is a small parameter near threshold.
Hence, the contribution of the ${\cal O}(\alpha_s^n)$ ladder diagram
will not be small even for a large $n$ if $\beta \simlt \alpha_s$.
These $(\alpha_s/\beta)^n$ 
singularities which appear at this specific kinematical
configuration are known as ``threshold singularities'' or
``Coulomb singularities''.

Intuitively the appearance
of $(\alpha_s/\beta)^n$ 
can be interpreted as follows.
When the produced $t$ and $\bar{t}$ 
have small velocities ($\beta \simlt \alpha_s$), they are trapped by the
attractive force mediated by the exchange of
gluons.
Thus, they stay close to each other for a long time and
multiple exchanges of gluons (higher order ladder diagrams)
become more significant
and the strong interaction is enhanced accordingly.

Since the higher order terms remain unsuppressed
in the threshold region,
we are led to resum these contributions.
The resummation technique is known since long time.
The leading $(\alpha_s/\beta)^n$ terms can be incorporated in 
the $t\bar{t}$ pair production vertex as\footnote{
See, for example, Ref.~\cite{sp} for a derivation of
the vertex $\Gamma^\mu$.
Also, Refs.\ \cite{dthesis,cracow} shows explicitly the appearances of
the $(\alpha_s/\beta)^n$ terms from all the ladder diagrams.
}
\bea
\Gamma^\mu = - \, \gamma^\mu 
\, ( E - {\bf p}_t^2/m_t + i\Gamma_t ) \,
\tilde{G}( {\bf p}_t;E ),
\label{ttvtx0}
\eea
where $E = \sqrt{s}-2m_t$ is the energy measured from the threshold.
$\tilde{G}( {\bf p};E )$ is the momentum-space Green function
of the non-relativistic Schr\"{o}dinger equation with the
Coulomb potential:
\bea
&&
\left[
\left( - \, \frac{\nabla^2}{m_t} + V(r) \right) 
- ( E + i\Gamma_t )
\right]
G({\bf x};E) = \delta^3({\bf x}),
\label{scheq0}
\\ && ~~~
\tilde{G}( {\bf p};E )
= \int d^3{\bf x} \, e^{-i{\bf p} \cdot {\bf x} }
\, G({\bf x};E) ,
\\ && ~~~~~~
V(r) = - \, C_F \, \frac{\alpha_s}{r} ,
\label{coulombpot}
\eea
where $C_F = 4/3$ is the color factor.

It is possible to perform a systematic perturbative
expansion of the cross sections
in the threshold region.
Roughly speaking, one may identify
\bea
\begin{array}{llc}
{\displaystyle \sum_{~} c_n^{(0)} (\alpha_s^n /\beta^n) }
&:& \mbox{leading} \\
{\displaystyle \sum_{~} c_n^{(1)} (\alpha_s^{n+1}/\beta^n) }
&:& {\cal O}(\alpha_s)~\mbox{correction} \\
{\displaystyle \sum_{~} c_n^{(2)} (\alpha_s^{n+2}/\beta^n) }
&:& {\cal O}(\alpha_s^2)~\mbox{correction} \\
~~~~~~~\vdots&&\vdots
\end{array}
\nonumber
\eea
For all interesting physical quantities (cross sections) for
$e^+e^- \to t\bar{t}$ near threshold, calculations
of the full
${\cal O}(\alpha_s)$ corrections have been completed so far
\cite{dthesis,ps}.
Meanwhile calculations of the second order corrections are
currently in progress.

One typical example of the ${\cal O}(\alpha_s)$ corrections is
the radiative corrections to the Coulomb-gluon-exchange kernel,
whose net effect is to replace the fixed coupling constant in
the Coulomb potential in Eq.~(\ref{coulombpot}) by the
running coupling constant,
$\alpha_s \to \alpha_s(\mu \simeq 1/r)$.
We thus have the QCD potential which becomes weaker
than the Coulomb potential at
short distances.

Recently, there 
has been considerable progress in the theoretical calculations
of the higher order corrections to the Coulomb bound-state problems.
New contributions have been calculated analytically for QED bound-states
\cite{hoang,hl},
which could not be achieved using the conventional bound-state approaches.
The corrections that originate from the relativistic regime and those from 
the non-relativistic regime have been separated using an effective Lagrangian
formalism \cite{cl}. 
The real difficult part of the calculations is now reduced to the
ordinary second-order (relativistic) perturbative calculation
of the cross section, which requires no knowledge of the bound-state problems.
The readers are referred to e.g.\ Refs.~\cite{labelle,gr} for 
introduction to the formalism.

\section{Production of Top Quark}

To understand the physics concerning the production of
top quarks, one needs to keep in mind that the $t\bar{t}$ production
vertex is proportional to the Green function of the
non-relativistic Schr\"odinger equation:
\bea
\Gamma^\mu \propto
\tilde{G}( {\bf p}_t;E ) =
- \sum_n
\frac{\phi_n (\vc{p}_t)\psi_n^* (\vc{0})}{E-E_n+i\Gamma_t} ,
\eea
where
\bea
\left[ \frac{\hat{\vc{p}}^2}{m_t} + V_{\rm QCD}(r) \right]
\ket{n} = E_n \ket{n} ,
~~~
\begin{array}{l}
\phi_n(\vc{p}) = \braket{\vc{p}}{n} \rule[-3mm]{0mm}{8mm}
\\
\psi_n(\vc{x}) = \braket{ \vc{x}}{ n} 
\end{array} 
\eea
defines the wave functions of the energy eigenstate
of the QCD potential.

\subsection{Total Cross Section}

\begin{figure}
\begin{center}
  \leavevmode
  \epsfxsize=7.cm
  \epsffile{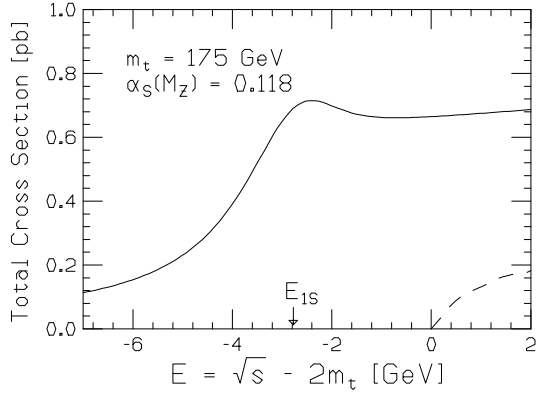}\vspace{3mm}\\  
  \caption[]{
        \label{totcs}The total cross section vs.\ energy, 
$E = \sqrt{s}-2m_t$.
The solid curve is calculated from the Green function.
The dashed curve shows the tree-level total cross section for
a stable top quark.
}
\end{center}
\end{figure}

The first observable we measure in the $t\bar{t}$ threshold region
will be the total cross section.
Via the optical theorem, the total cross section can be written as
\cite{fk,sp}
\bea
\sigma_{\rm tot} (e^+e^- \to t\bar{t})
\propto
- {\rm Im}
\sum_n
\frac{|\psi_n(\vc{0})|^2}{E-E_n+i\Gamma_t} .
\eea
One sees that the energy dependence of the total
cross section is determined by the resonance spectra.
Due to the large width $\Gamma_t$
of the top quark, however, distinct resonance peaks are smeared out.
The resonances merge with one another, leading to a
broad enhancement of the cross section over the threshold region
as seen in Fig.~\ref{totcs}.
(We will show explicitly the resonance spectra below.)
In the same figure, the tree level cross section is also shown as a dashed
curve.
Despite the disappearance of each resonance peak, one sees that 
the cross section is indeed largely enhanced by the QCD interaction,
and that inclusion of
the QCD binding effect is mandatory for a proper account of the
cross section in the threshold region.

\subsection{Top Momentum Distribution}

\begin{figure}
\begin{center}
  \leavevmode
  \epsfxsize=5.5cm 
  \mbox{\epsffile[82  394  505  749]{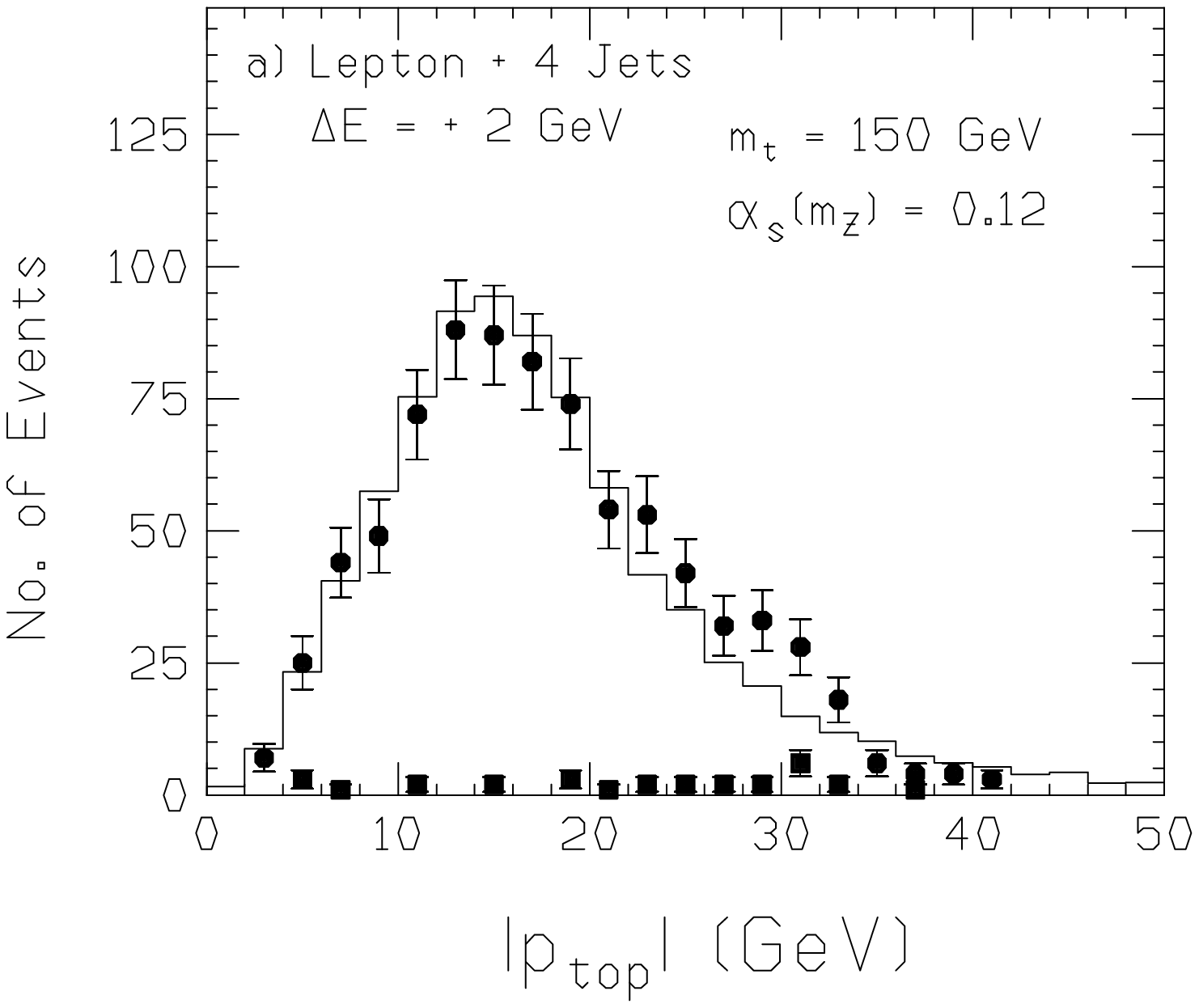}}\hspace{5mm}
  \raise2mm\hbox{\epsfxsize=5.5cm \epsffile{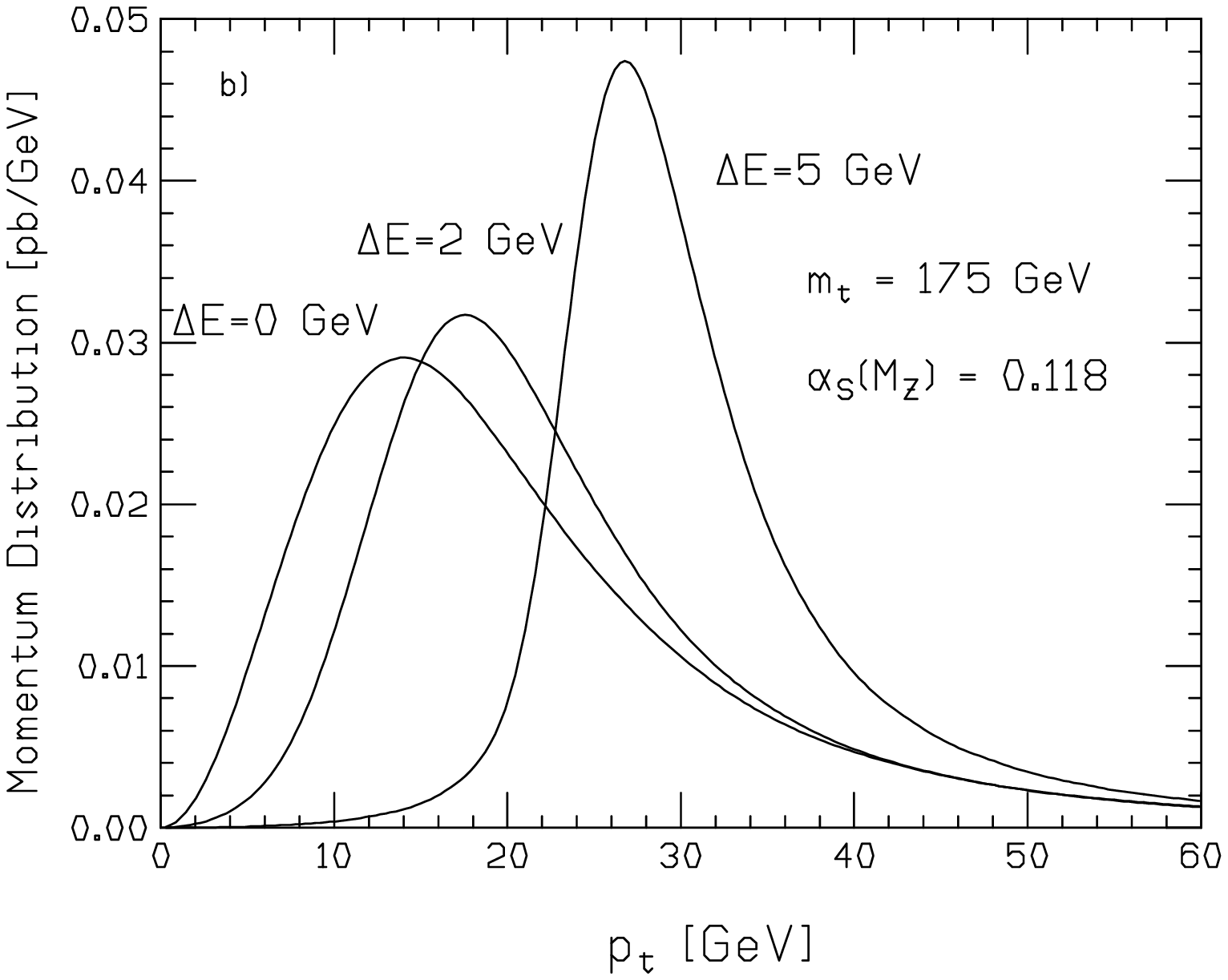}}\\  
\end{center}\vspace{-2mm}
  \caption[]{
        \label{momdist}
(a) Reconstructed momentum distribution (solid circles) for the 
lepton-plus-4-jet mode, compared with the generated distribution (histogram).
The Monte Carlo events were generated with 
$\alpha_s(M_Z) = 0.12$ and $m_t = 150$~GeV \cite{fms}.
(b) Top-quark momentum distributions $d\sigma/d|\vc{p}_t|$
for various c.m.\ energies measured from
the lowest lying resonance, $\Delta E = \sqrt{s} - M_{1S}$,
taking $\alpha_s(M_Z) = 0.118$ and $m_t = 175$~GeV.
}
\end{figure}

Next we consider the top-quark momentum ($|\vc{p}_t|$) distribution
near $t\bar{t}$ threshold \cite{sfhmn,jkt}.
It has been shown that experimentally it will be possible to reconstruct
the top-quark momentum $\vc{p}_t$ from its decay products with
reasonable resolution and detection efficiency.
Fig.~\ref{momdist}(a) 
shows a comparison of reconstructed top momenta (solid circles)
with that of generated ones (histogram), 
where the events are generated by a Monte Carlo generator
and are reconstructed after going through detector simulators
and selection cuts \cite{fms}.
The figure demonstrates that the agreement is fairly good.

Theoretically, the top-quark momentum distribution is given by
\bea
\frac{d\sigma}{d|\vc{p}_t|} &\propto&
\biggl| \sum_n
\frac{\phi_n (\vc{p}_t)\psi_n^* (\vc{0})}{E-E_n+i\Gamma_t}
\biggl|^2  + \mbox{(sub-leading)} .
\label{msgf}
\eea
The $|\vc{p}_t|$-distribution is thus governed by the momentum-space
wave functions of the resonances. 
By measuring the momentum distribution, essentially we measure 
(a superposition of) the wave functions of the toponium resonances.
Shown in Fig.~\ref{momdist}(b) are the top momentum distributions for various
energies.
One may also vary the magnitude of $\alpha_s$ and confirm that the
distribution is indeed sensitive to the resonance wave functions
\cite{sfhmn,jkt}.
Hence, the momentum distribution provides information independent of that
from the total cross section.

Note that the toponium states will be the first quarkonium resonances whose
wave functions can be measured experimentally.
For comparison, consider $\Upsilon (4S)$, 
which decays into $B$ and $\bar{B}$.
Since the mass of $B$($\bar{B}$) is fixed, its momentum is
fixed by the on-shell condition; the momentum distribution 
of $B$($\bar{B}$) is a $\delta$-function in this case.
Meanwhile,
in the case of toponium, the invariant mass distribution of
top quarks has a large width, and accordingly the top-quark
three momenta have a distribution that
is just sufficiently broad for probing the wave functions of the
resonances.

\subsection{Forward-Backward Asymmetry}

Another observable that can be measured experimentally
is the forward-backward asymmetry of the top quark \cite{ms}.
Generally in a fermion pair production process, a forward-backward asymmetric
distribution originates from an interference of
the vector and axial-vector $f\bar{f}$ production vertices at tree level
of electroweak interaction.
One can show from the spin-parity
argument that in the threshold region the $t\bar{t}$ vector vertex 
creates S-wave resonance states, while the 
$t\bar{t}Z$ axial-vector vertex creates P-wave states.
Therefore, by observing the forward-backward asymmetry of the top quark,
we observe an interference of the S-wave and P-wave states.

\begin{figure}
\begin{center}
  \leavevmode
  \epsfxsize=7.cm
  \epsffile{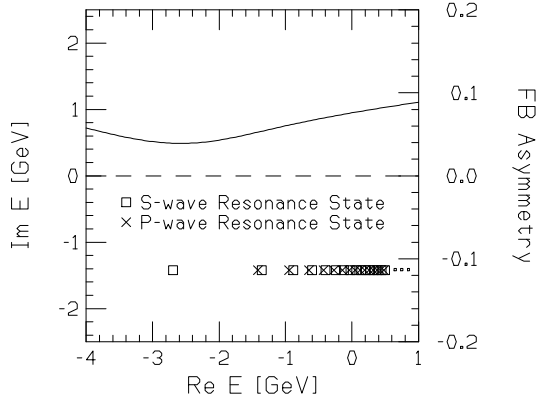}\vspace{3mm}\\  
  \caption[]{
        \label{fbasym}The positions of poles of the S-wave and P-wave
states on the complex energy plane, together with the forward-backward
asymmetry as a function of the energy, taking
$\alpha_s(M_Z) = 0.118$ and $m_t = 175$~GeV.
The right-axis is for the forward-backward asymmetry.
}
\end{center}
\end{figure}

In general, S-wave resonance states and P-wave resonance states have
different energy spectra.
So if the c.m.\ energy is fixed at either of the spectra, there would be no
contribution from the other.
However, the widths of resonances are large for the toponium
in comparison to their level splittings,
which permit sizable interferences of the S-wave and P-wave 
states.\footnote{
Note that no forward-backward asymmetry is observed for charmonium or
bottomonium states because the widths of the resonances are too small
compared to their level splittings.
Thus, the asymmetry 
reveals to be another observable unique to the toponium states.
}
Fig.~\ref{fbasym} 
shows the pole position $E_n \! - \! i\Gamma_t$ of these states
on the complex energy plane.
One sees that the widths of the resonances are comparable to the
mass difference between the lowest lying S-wave and P-wave states, and
exceeds by far the level spacings between higher S-wave and P-wave states.
This gives rise to a forward-backward 
asymmetry even below threshold, and provides
information on the resonance level structure which is concealed in the total
cross section.
Shown on the same figure is the forward-backward asymmetry as a function
of the energy.
It is seen that the asymmetry takes its minimum value
at around the lowest lying S-wave state, where the interference is
smallest, and
increases up to $\sim 10$\% with energy as the resonance spectra appear
closer to each other.
One may also confirm that essentially the forward-backward asymmetry 
measures the degree of overlap of the S-wave and P-wave states
by varying the coupling constant $\alpha_s$ or the top quark
decay width $\Gamma_t$.

The resonance level 
structure is determined by QCD, whereas the resonance widths are
determined by electroweak interaction.
The interplay of the two interactions generates the forward-backward 
asymmetry.

\section{Decay of Top Quark and Final-State Interactions}

Now we turn to decay processes of the top quark near threshold.
The top quarks produced via $e^+e^- \to t\bar{t}$ in the
threshold region will be highly polarized \cite{kuehn}.
Even for an unpolarized $e^-$ beam, the top quarks have a natural
polarization of around 40\%, while for a longitudinally
polarized $e^-$ beam (an obvious option
for NLC) the polarization of top quarks can be raised close to
100\% \cite{r38,r26}.
Therefore, in principle, the threshold region can be an ideal place
for studying the top quark decay processes using the highly polarized
top quark samples and the largest $t\bar{t}$ production cross section.

\subsection{Free Polarized Top Quark Decay}

Detailed studies of the decay of {\it free} polarized top quarks have
already been available including the full ${\cal O}(\alpha_s)$ corrections
\cite{r44,r27,r28}.
A nice example is that of the energy-angular distribution of charged
leptons $l^+$ in the semi-leptonic decay of the top quark.
In leading order, the $l^+$ distribution
has a form where the energy and angular dependences are factorized
\cite{ks,r43}:
\bea
\frac{d\Gamma_{t \to bl^+\nu}
(\vc{S})}
{dE_l d\Omega_l} 
= h(E_l) \, ( 1 + |\vc{S}| \cos \theta_l ) \, + \,
({\cal O}(\alpha_s)~\mbox{correction}) .
\eea
Here $E_l$, $\Omega_l$, and $\theta_l$
denote, respectively, the $l^+$ energy, the solid angle of $l^+$, 
and the angle between the $l^+$ direction and the 
top polarization vector $\vc{S}$, all of which 
are defined in the top-quark rest frame.
Hence, we may measure the top-quark polarization 
with maximal sensitivity using the $l^+$ angular
distribution.

\subsection{Effects of Final-State Interactions}

\begin{figure}
\begin{center}
  \leavevmode
  \epsfxsize=10.cm
  \epsffile{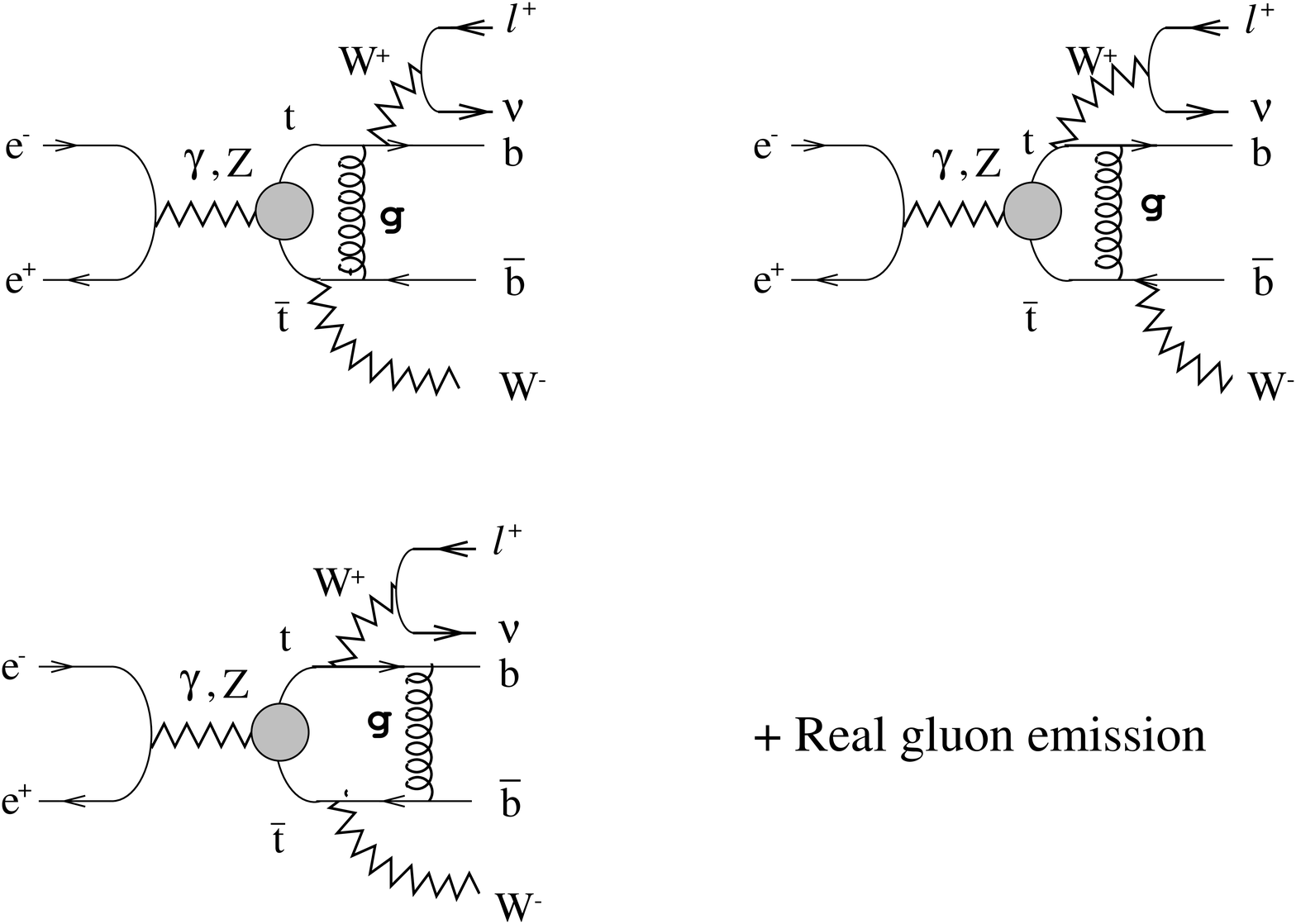}\vspace{3mm}\\  
  \caption[]{
        \label{fsidiagram}
Diagrams for the final-state interactions for 
$e^+e^- \to t\bar{t} \to bl^+\nu\bar{b}W^-$.
}
\end{center}
\end{figure}

Close to threshold, the above precise
analyses of the free top-quark decays do not apply directly
because of the existence of corrections unique to this region.
Namely, these are 
the final-state interactions due to gluon exchange between $t$
and $\bar{b}$ ($\bar{t}$ and $b$) or between $b$ and $\bar{b}$.
(Fig.~\ref{fsidiagram})\
The size of the corrections is at the 10\% level 
in the threshold region, hence it is 
necessary to incorporate their
effects in precision studies of top-quark production and decay near
threshold.

Before presenting the formula for these
final-state interaction corrections, let us first see what
kind of effects we expect from physics ground \cite{ps}.
\\

\noindent
$\bullet$ {\it Top Momentum Distribution}
\medbreak

\begin{figure}
\begin{center}
  \leavevmode
  \epsfxsize=6.cm \epsffile{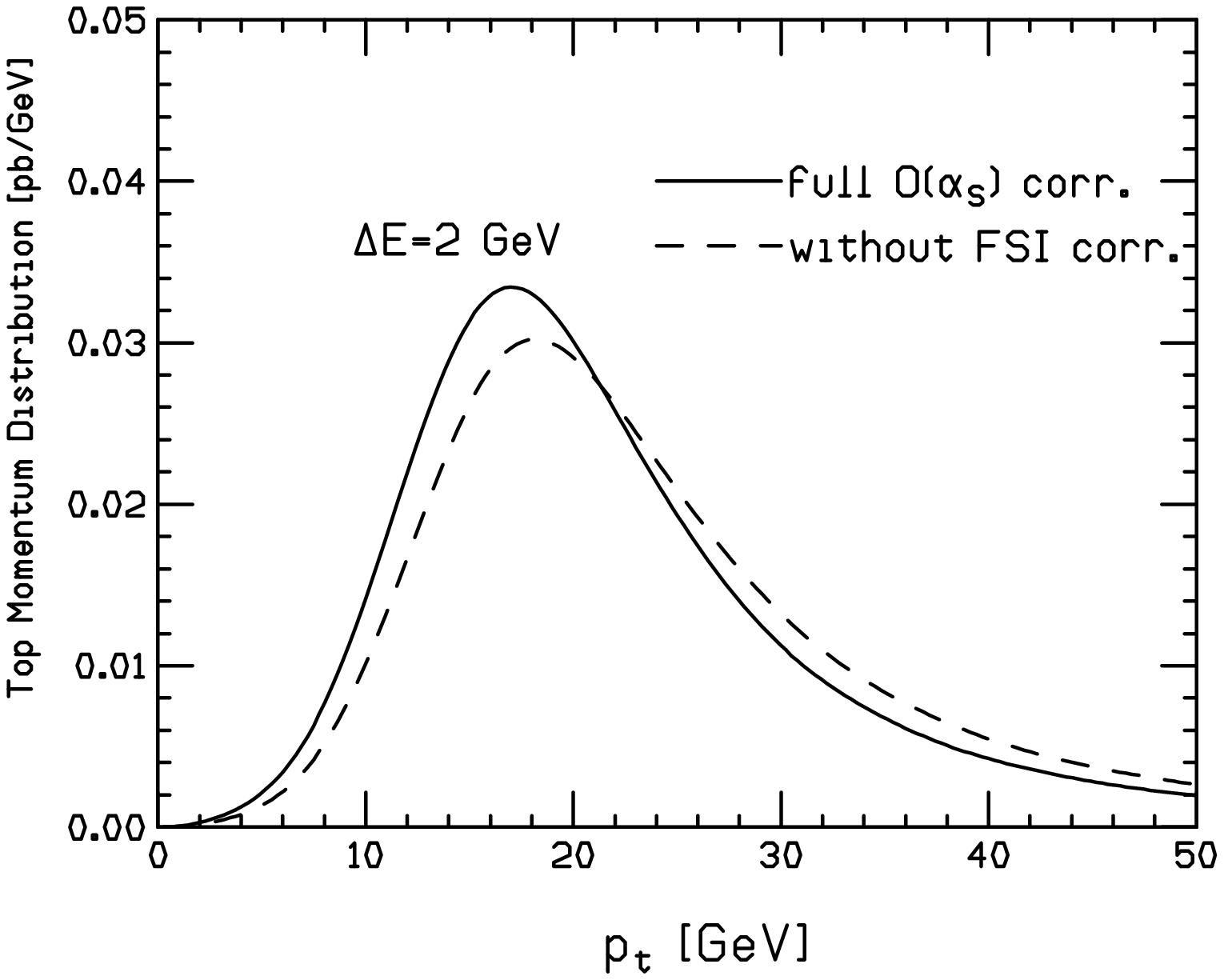}\hspace{8mm} 
  \raise7mm\hbox{\epsfxsize=4.3cm \epsffile{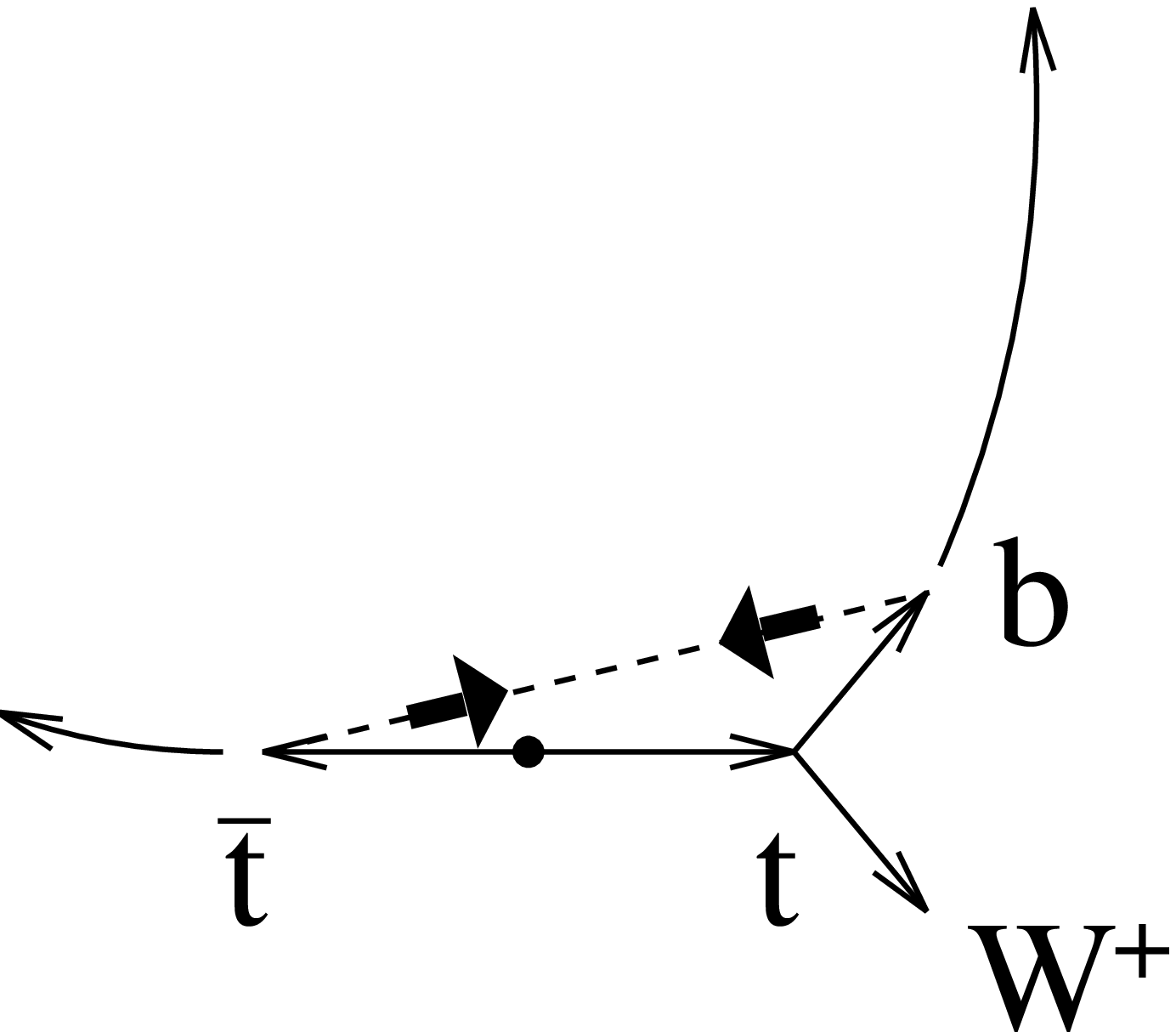}}
  \vspace{1mm}\\  
  \end{center}
  \caption[]{
        \label{mdistfsi}
(a) The top momentum distribution with (solid) and
without (dashed) the final-state interaction corrections for
$\alpha_s(M_Z) = 0.118$ and $m_t = 175$~GeV.
(b) Attractive force between $\bar{t}$
and $b$ (from $t$ decay). The momentum transfer 
$\delta \vc{p}_b = - \delta \vc{p}_{\bar{t}}$ due to
the attraction is indicated by thick arrows.
}
\end{figure}

Perhaps it is easiest to understand
the effect of final-state interactions on the top
momentum distribution.
Fig.~\ref{mdistfsi}(a)
shows the momentum distribution with (solid) and without (dashed)
the final-state interactions.
We see that the average momentum is reduced due to the interaction.
To understand this,
consider for example the case where $t$ decays first.
Fig.~\ref{mdistfsi}(b) shows the attractive force between
$\bar{t}$ and $b$, which deflects the trajectory of $b$.
Since $\vc{p}_t$ is reconstructed from the $bW^+$ momenta at time
$\tau \to \infty$, it is obvious that
the reconstructed momentum 
$|\vc{p}_t| = |\vc{p}_b + \vc{p}_{W^+}|$ is decreased by the 
attraction.
\\

\noindent
$\bullet$ {\it Forward-Backward Asymmetric Distribution}
\medbreak

\begin{figure}
\begin{center}
  \leavevmode
  \epsfxsize=4.6cm
  \epsffile{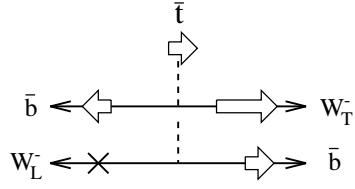}\vspace{3mm}\\ 
  \caption[]{
        \label{tbardecay}Typical configurations in the decay 
        of $\bar{t}$ with definite
        spin orientation.  Transverse $W^-$ ($W^-_T$) tends 
        to be emitted in the direction of the $\bar{t}$ spin orientation,
        while longitudinal $W^-$ ($W^-_L$) is emitted 
        in the opposite direction due to helicity conservation.
        For $m_t \simeq 175$~GeV, $\bar{t}$ decays mainly to
        $W^-_L$, hence $\bar{b}$ is emitted more in the $\bar{t}$ spin
        direction.}
\end{center}
\end{figure}

\begin{figure}
\begin{center}
  \leavevmode
  \epsfxsize=9.cm
  \epsffile{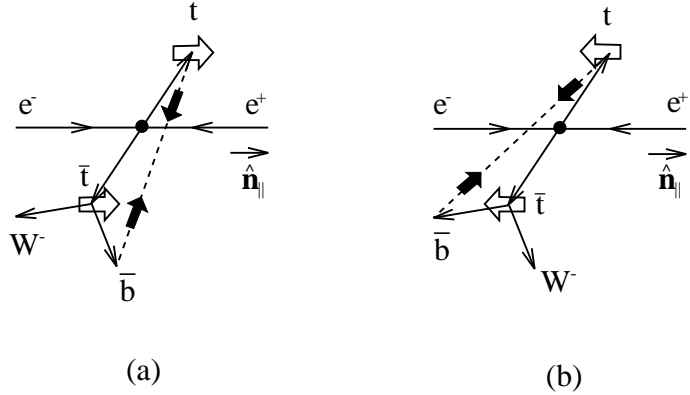}\vspace{3mm}\\
  \caption[]{
        \label{spin}Attractive force between $t$ and $\bar{b}$
        when the $t$ and $\bar{t}$ spins are oriented in the
        (a) $\hat{\vc{n}}_\|$ direction, and in the
        (b) $-\hat{\vc{n}}_\|$ direction. The 
        momentum transfer  
        $\delta \vc{p}_b = - \delta \vc{p}_{\bar{t}}$ due to
        the attraction is indicated by thick black arrows.}
\end{center}
\end{figure}

Next we consider the $\cos \theta_{te}$ distribution of the top quark.
($\theta_{te}$ denotes the angle between $t$ and $e^-$ 
in the $t\bar{t}$ c.m.\ frame.)
We consider the case where $\bar{t}$ decays first and examine the
interaction between $t$ and $\bar{b}$.
The $t$ and $\bar{t}$ pair-produced near threshold in $e^+e^-$ collisions
have their spins approximately parallel or anti-parallel to the $e^-$ beam
direction ($\hat{\vc{n}}_\|$) and
the spins are always oriented parallel to each other.
On the other hand, the decay of $\bar{t}$ occurs via a $V \! -\! A$
coupling, and
$\bar{b}$ is emitted preferably in the spin direction of the parent $\bar{t}$,
see Fig.\ \ref{tbardecay}.
More precisely,
the excess of the $\bar{b}$'s 
emitted in the $\bar{t}$ spin direction over those
emitted in the opposite direction is given by 
$\kappa = (m_t^2 - 2M_W^2)/(m_t^2+2M_W^2)$.
Now suppose $t$ and $\bar{t}$ have their spins in the $\hat{\vc{n}}_\|$
direction.
Then $\bar{b}$ will be emitted dominantly in the $\hat{\vc{n}}_\|$
direction.
One can see from Fig.~\ref{spin}(a)
that in this case $t$ is always attracted to the forward direction 
due to the attractive force between $t$ and $\bar{b}$.
The direction of the attractive force will be opposite
if $t$ and $\bar{t}$ have
their spins in the $-\hat{\vc{n}}_\|$ direction (Fig.~\ref{spin}(b)).
Thus, polarized top quarks will be pulled in a
definite (forward or backward) direction,
and we may expect 
that a forward-backward asymmetric distribution of
the top quark
$\sim \kappa \, S_\| \cos \theta_{te}$ is
generated by the final-state interaction.
($S_\|$ denotes the $\hat{\vc{n}}_\|$-component of the 
top polarization vector $\vc{S}$.)
\\

\noindent
$\bullet$ {\it Top-Quark Polarization Vector}
\medbreak

From Fig.\ \ref{spin}
we can also learn the effect of the final-state interaction
on the top polarization vector.
We have seen
that if the $t$ and $\bar{t}$ spins are oriented in the $\hat{\vc{n}}_\|$ 
direction, $t$ will be attracted to the forward direction
due to the attraction by $\bar{b}$, and oppositely attracted
to the backward direction if the $t$ and $\bar{t}$ spins are 
in the $-\hat{\vc{n}}_\|$ direction.
This means that in the forward region ($\cos \theta_{te} \simeq 1$)
the number of $t$'s with
spin in $\hat{\vc{n}}_\|$ direction increases, whereas in the
backward region the number of 
those with spin in the opposite direction increases.
Or equivalently, the $\hat{\vc{n}}_\|$-component of
the top-quark polarization vector
increases in the forward region and decreases in the backward region.
We may thus conjecture that the top-quark polarization vector is modified
as 
$\delta S_\| \sim \kappa \cos \theta_{te}$
due to the interaction between $t$ and $\bar{b}$.
\\

\noindent
$\bullet$ {\it $l^+$ Energy-Angular Distribution}
\medbreak

\begin{figure}
\begin{center}
  \leavevmode
  \epsfxsize=10.cm
  \epsffile{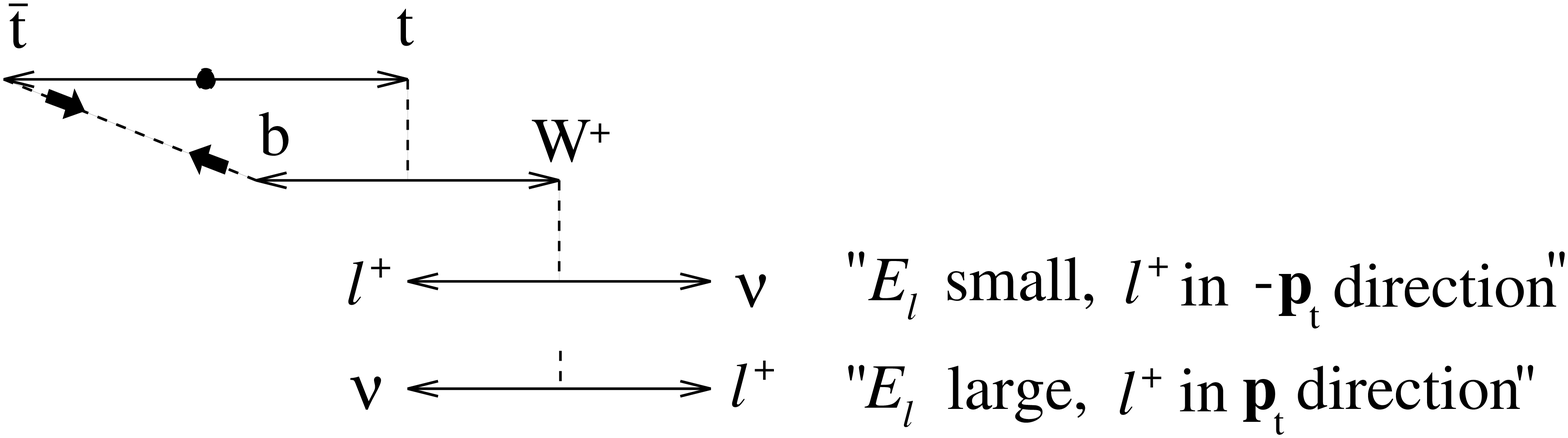}\vspace{3mm}\\
  \caption[]{
        \label{semildecay}Typical configurations of the particles
        in semileptonic decay of $t$ when the $b$-quark
        is emitted in the $\bar{t}$ direction.
        Due to the boost by $W^+$, there will be an
        energy-angle correlation of $l^+$.
}
\end{center}
\end{figure}

Finally let us examine the effect of the attraction between
$b$ and $\bar{t}$ on the $l^+$ energy-angular distribution in the
semi-leptonic decay of $t$.
The $b$-quark from $t$ decay
will be attracted in the direction of $\bar{t}$ due to
the gluon exchange between these two particles.
We show schematically typical configurations of the particles
in the top-quark semileptonic decay in Fig.\ \ref{semildecay}.
It can be seen that if the probability for $b$ being emitted in
the $\bar{t}$ direction increases, correspondingly
the probability for particular $l^+$ energy-angular configurations 
increases.
These configurations are either
``$E_l$ is small and $l^+$ emitted in $- \vc{p}_t$ direction'' or
``$E_l$ is large and $l^+$ emitted in $\vc{p}_t$ direction''.

\subsection{Lepton Energy-Angular Distribution Near $t\bar{t}$ Threshold}

Here, we present the formula for the charged 
lepton energy-angular distribution
in the decay of top quarks that are produced via $e^+e^- \to t\bar{t}$
near threshold.

First, without including the final-state interactions,
the differential distribution of $t$ and $l^+$ has a form where the
production and decay processes of the top quark
are factorized \cite{grzadkowski}:
\bea
\frac{d\sigma(e^+e^- \! \to t\bar{t} \to bl^+\nu\bar{b}W^-)}
{d^3\vc{p}_t dE_l d\Omega_l}
=
\frac{d\sigma(e^+e^- \! \to t\bar{t})}
{d^3\vc{p}_t} \times
\frac{1}{\Gamma_t} \,
\frac{d\Gamma_{t \to bl^+\nu}
(\vc{S}) }
{dE_l d\Omega_l} .
\label{born}
\eea
Namely, the cross section is given as a product of the production
cross section of unpolarized top quarks and the differential decay
distribution of $l^+$ from polarized top quarks.
The above formula holds true even including all
${\cal O}(\alpha_s)$ corrections other than the final-state
interactions.

Including the final-state interactions,
the factorization of production and decay processes is destroyed.
The formula including the full ${\cal O}(\alpha_s)$ corrections
is given by \cite{ps}
\bea
\frac{d\sigma(e^+e^- \! \to t\bar{t} \to bl^+\nu\bar{b}W^-)}
{d^3\vc{p}_t dE_l d\Omega_l}
=
\frac{d\sigma(e^+e^- \! \to t\bar{t})}
{d^3\vc{p}_t} \times
( 1 + \delta_0 + \delta_1 \cos \theta_{te} )
\nonumber \\
\times
\frac{1}{\Gamma_t} \,
\frac{d\Gamma_{t \to bl^+\nu}
(\vc{S}+\delta \vc{S}) }
{dE_l d\Omega_l} 
\times \left[ 1 +
\xi (|\vc{p}_t|,E,E_l,\cos\theta_{lt}) \right] .
\nonumber \\
\label{full}
\eea
Here, the first line on the right-hand-side shows that 
there are corrections to the top-quark
production cross section,
while the second line shows that the correction to the 
decay distribution of $l^+$ is
accounted for by a modification of the parent top-quark polarization vector,
and finally there is a non-factorizable correction $\xi$ which cannot be
assigned either to the production or the decay process alone.

We have already seen in Fig.~\ref{mdistfsi}(a) that 
the top momentum distribution
is modified by $\delta_0$ to take a lower average momentum.
The forward-backward asymmetric distribution 
and the top polarization
vector get corrections as
\bea
&&
\delta_1 \cos \theta_{te} =
\kappa S_\| \cos \theta_{te} \times \frac{1}{2}\psi_{\rm _R} ,
\label{del1}
\\
&&
\delta \vc{S} =
\left[ 1 - (S_\|)^2 \right] \times
\kappa \cos \theta_{te} \times {\frac{1}{2}} \psi_{\rm _R}
\cdot \hat{\vc{n}}_\| 
\label{delS}
\eea
with
\bea
&&
\psi_{\rm _R}(|\vc{p}_t|,E) = - \, C_F \! \cdot \! 4 \pi \alpha_s \,
\mbox{Pr.}\int \mbox{$\frac{d^3\vc{q}}{(2\pi)^3}$} \,
\, \frac{1 \,}{|\vc{q}\! -\! \vc{p}_t|^3} \,
\frac{\vc{p}_t\! \cdot \! (\vc{q}\! -\! \vc{p}_t)}{|\vc{p}_t|\, 
|\vc{q}\! -\! \vc{p}_t|}
\,
2 \,
\mbox{Re}
\biggl[ \frac{\tilde{G}(\vc{q};E)}{\tilde{G}(\vc{p}_t;E)} \biggl] .
\nonumber \\
\eea
The above formulas (\ref{del1}) and (\ref{delS}) have exactly 
the forms that we 
anticipated in the previous subsection if
$\psi_{\rm _R}$ is positive.
Indeed, the numerical evaluation in Ref.\ \cite{r26} shows that 
$\psi_{\rm _R}(|\vc{p}_t|,E) \simgt 0$ holds in the entire 
threshold region.\footnote{
It shows that the force between $b$ and $\bar{t}$ ($\bar{b}$ and $t$)
is attractive in the entire threshold region.
Note that the sign of $\psi_{\rm _R}$ will be reversed if the 
force is repulsive.
}

\begin{figure}
\begin{center}
  \leavevmode
  \epsfxsize=7.cm
  \epsffile{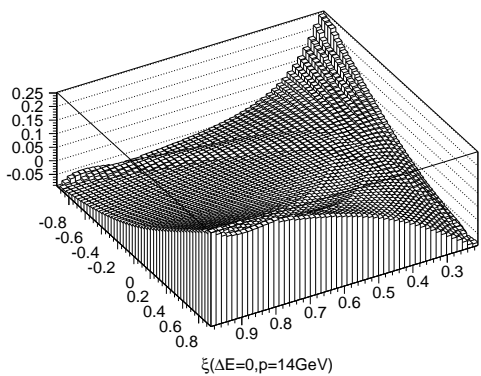}\vspace{3mm}\\
  \caption[]{
        \label{xi}
A three-dimensional plot of $\xi$ as a function of $2E_l/m_t$ ($x$-axis) and
$\cos\theta_{te}$ ($y$-axis).
}
\end{center}
\end{figure}

We show the $\cos \theta_{lt}$ and $E_l$ dependences of the
non-factorizable correction $\xi$ as a 3-dimensional plot in
Fig.~\ref{xi}.
One can see that $\xi$ takes comparatively large
positive values for
either ``small $E_l$ and $\cos \theta_{lt} \simeq -1$'' or
``large $E_l$ and $\cos \theta_{lt} \simeq +1$''.
Oppositely,
in the other two corners of the $E_l$--$\cos \theta_{lt}$ plane
$\xi$ becomes negative.
These features are consistent
with our previous qualitative argument.
The typical magnitude of $\xi$ is 10--20\%.

Thus, the theoretical prediction for the distribution of $l^+$ from 
the top decay is
under good control in the $t\bar{t}$ threshold region, 
together with a good qualitative understanding.

Prior to the calculation of the $l^+$ differential distribution
Eq.~(\ref{full}),
an inclusive quantity, the mean value $\langle n \ell \rangle$
of the charged lepton four-momentum projection on an arbitrarily chosen
four-vector $n$, was proposed as an observable sensitive
to the top quark polarization, and this quantity was calculated
including the final-state interactions
\cite{r26}.

\subsection{Observable Proper to the Decay Process}

We have seen that the final-state interactions
destroy the factorization of the 
production and decay cross sections of the top quark.
In order to study the decay of the top quark in a clean environment in
the threshold region, 
it would be useful if we could find an observable which depends
only on the decay process
of free polarized top quarks, 
$d\Gamma_{t \to bl^+\nu}(\vc{S})/ dE_l d\Omega_l$.
In fact, such an observable can be constructed, which
at the same time preserves most of the 
differential information of the $l^+$ energy-angular distribution.

It is possible to show (with sufficient reasoning)
that the non-factorizable correction factor
$\xi (|\vc{p}_t|,E,E_l,\cos\theta_{lt})$
is invariant under a transformation of the $l^+$ kinematical variables
\bea
E_l &\to& E'_l = 2 m_t
\left( \frac{m_t^2 + M_W^2}{M_W^2}- \frac{2m_t}{E_l} \right) ^{-1} ,
\\
\vc{p}_l &\to& \vc{p}_l'
~~~~\mbox{such that}~~~
\cos\theta_{lt} \to -\cos\theta_{lt} .
\eea
Using this symmetry, it is possible to
cancel out not only the non-factorizable correction but also
the top production cross section by taking an appropriate ratio of
cross sections.

Let us define an observable as
\bea
&&
\overline{A}( E_l, a \! = \! \vc{S} \! \cdot \! \vc{p}_l )
\nonumber \\
&& 
\equiv
\frac{\displaystyle
\int {d^3\vc{p}_t d\Omega_l}
\, \, \delta \!
\left( \vc{S} \! \cdot \! \vc{p}_l - a \right)
\left[
\frac{d\sigma(e^+e^- \! \to t\bar{t} \to bl^+\nu\bar{b}W^-)}
{d^3\vc{p}_t dE_l d\Omega_l}
\right]_{E_l,\vc{p}_l}
}{\displaystyle
\int {d^3\vc{p}_t d\Omega_l}
\, \, \delta \!
\left( \vc{S} \! \cdot \! \vc{p}_l  + a \right)
\left[
\frac{d\sigma(e^+e^- \! \to t\bar{t} \to bl^+\nu\bar{b}W^-)}
{d^3\vc{p}_t dE_l d\Omega_l}
\right]_{E'_l,\vc{p}_l}
} .
\label{def}
\eea
Here, the top-quark polarization vector $\vc{S}$ in the delta functions 
depends on $\vc{p}_t$ \cite{ps}.
The numerator and denominator, respectively, depend 
on two external kinematical variables
(the lepton energy and the lepton angle from the
parent top-quark polarization vector), and all other 
variables are integrated out before taking the ratio.

Then substituting the differential distribution Eq.~(\ref{full}),
one can show that theoretically $\overline{A}$ is determined solely 
from the decay distribution of free polarized top quarks:
\bea
\overline{A}( E_l, a )
=
\left[ \frac{d\Gamma_{t \to bl^+\nu}(\vc{S}) }
{dE_l d\Omega_l} 
\right]_{E_l, \vc{S} \cdot \vc{p}_l = a} 
\biggl/
\left[
\frac{d\Gamma_{t \to bl^+\nu}(\vc{S}) }
{dE_l d\Omega_l} 
\right]_{E_l', \vc{S} \cdot \vc{p}_l = - a} 
.
\label{thabar}
\eea
This is a general formula that is
valid even if the decay vertices of the top quark deviate
from the standard-model forms.

This quantity will be useful from the theoretical point of view.
If one claims that he calculates 
$\overline{A} ( E_l, a )$ defined in Eq.\ (\ref{def})
in the threshold region, it can be calculated without including any
bound-state effects or final-state interaction corrections but
only from a decay distribution of free polarized top quarks
via Eq.~(\ref{thabar}).

\section{Summary}

We have reviewed the physics concerning the production and decay of
top quarks in the $t\bar{t}$ threshold region.

Theoretical predictions of cross sections near $t\bar{t}$ threshold are well
under control.
The full ${\cal O}(\alpha_s)$ corrections as well as part
of the second order corrections are already available.

As for the top-quark production process, 
there are three independent observables that are unique to
the $t\bar{t}$ threshold region.
The total cross section is enhanced by the QCD interaction,
but distinct resonance peaks are smeared out 
due to the large decay widths of the resonances.
The top quark momentum measurement will probe
the resonance wave functions.
The top-quark forward-backward asymmetry 
measures the overlaps of the S and P-wave resonance states.

Studies of the decay of top quarks in the $t\bar{t}$ threshold region have
just been started.
First, an inclusive observable $\langle n \ell \rangle$
was calculated, which is sensitive to the top polarization.
Recently, the differential decay distribution
of $l^+$ in the top-quark semileptonic decay has been calculated.
The final-state interactions 
modify the top-quark production cross section,
the top-quark polarization vector, and also gives rise to a non-factorizable
correction at the level of 10--20\%.

We defined a new observable $\overline{A} ( E_l, a )$
in the threshold region, 
which depends only on the decay process
of free polarized top quarks.
This quantity can be calculated (including e.g.\ anomalous top-quark 
decay vertices)
without any knowledge of the bound-state
effects or the final-state interactions, but assuming the highly polarized top
quark samples expected in the $t\bar{t}$ threshold region.
Further studies in this direction are demanded.

Finally, a supplementary remark would be in order
for those who are interested to know how accurately
various physical parameters 
($m_t$, $\alpha_s$, $\Gamma_t$, $M_H$, $g_{tH}$, etc.)\
can be measured in the $t\bar{t}$ threshold region at NLC.
The results from quantitative studies 
taking into account realistic experimental conditions 
can be found in Refs.\ \cite{fms,cracow,cmmo}
for $m_t = 150$~GeV, 170~GeV and 180~GeV, respectively.
\medbreak

The first half of the paper is based on the studies 
in collaboration with K.~Fujii, K.~Hagiwara, K.~Hikasa, S.~Ishihara, 
T.~Matsui, H.~Murayama and C.-K.~Ng.
The latter half is based on our recent work with M.~Peter.
The author wishes to thank all of them.
The author is also grateful to A.~Hoang,
M.~Je\.{z}abek and J.~K\"{u}hn for fruitful discussion.
The author is thankful to K.~Melnikov, M.~Peter and S.~Recksiegel
for comments on the manuscript.
This work is supported by the Alexander von Humboldt Foundation.

\def\app#1#2#3{{\it Acta~Phys.~Polonica~}{\bf B #1} (#2) #3}
\def\apa#1#2#3{{\it Acta Physica Austriaca~}{\bf#1} (#2) #3}
\def\npb#1#2#3{{\it Nucl.~Phys.~}{\bf B #1} (#2) #3}
\def\plb#1#2#3{{\it Phys.~Lett.~}{\bf B #1} (#2) #3}
\def\prd#1#2#3{{\it Phys.~Rev.~}{\bf D #1} (#2) #3}
\def\pR#1#2#3{{\it Phys.~Rev.~}{\bf #1} (#2) #3}
\def\prl#1#2#3{{\it Phys.~Rev.~Lett.~}{\bf #1} (#2) #3}
\def\sovnp#1#2#3{{\it Sov.~J.~Nucl.~Phys.~}{\bf #1} (#2) #3}
\def\yadfiz#1#2#3{{\it Yad.~Fiz.~}{\bf #1} (#2) #3}
\def\jetp#1#2#3{{\it JETP~Lett.~}{\bf #1} (#2) #3}
\def\zpc#1#2#3{{\it Z.~Phys.~}{\bf C #1} (#2) #3}


\begin{thebibliography}{99}

\bibitem{nlc}
For reviews on physics that can be covered at NLC, see
{\it Proceedings of Workshop on 
Physics and Experiments with Linear Colliders}, Saariselka,
Finland, 1991, edited by R. Orava, P. Elrola and M. Nordbey (World
Scientific,
Singapore, 1992);
JLC Group, KEK Report 92-16 (1992);
W. Bernreuther, et al., 
in {\it $e^+e^-$ Collisions at 500 GeV: The Physics Potential, Part A, B, C
and D},
DESY 92-123, (1992), DESY 93-123, (1993), DESY 96-123, (1996);
{\it Proceedings of the Workshop on Physics and Experiments with 
Linear $e^+e^-$ Colliders} (Waikoloa, Hawaii, April 1993);
K. Fujii,
Talk given at {\it 22nd INS International Symposium on Physics with High
Energy Colliders},
KEK preprint, 94-38 (1994).


\bibitem{fk} V.S. Fadin and V.A. Khoze, {\it JETP ~Lett.}
{\bf 46}, 525 (1987);
{\it Sov.~J.~Nucl.~Phys.} {\bf 48}, 309 (1988).

\bibitem{komamiya} S. Komamiya,
in {\it Research Directions for the Decade, Proceedings of the 1990 
Summer Study
on High Energy Physics}, Snowmass, Colorado, 1990, edited by E. Berger (World
Scientific, Singapre, 1992), p.459.

\bibitem{fy} V. Fadin and Yakovlev,
{\it Sov.~J.~Nucl.~Phys.} {\bf 53}, 688 (1991).

\bibitem{kw} W. Kwong,
{\it Phys.~Rev.} {\bf D43}, 1488 (1991).

\bibitem{sp} M. Strassler and M. Peskin, 
{\it Phys.~Rev.} {\bf D43}, 1500 (1991).

\bibitem{peskin} M.~Peskin, in
{\it Proceedings of Workshop on 
Physics and Experiments with Linear Colliders}, Saariselka,
Finland, 1991, edited by R. Orava, P. Elrola and M. Nordbey (World
Scientific, Singapore, 1992).

\bibitem{sfhmn} Y. Sumino, K. Fujii, K. Hagiwara, H. Murayama, and
C.-K. Ng,  {\it Phys.~Rev.} {\bf D47}, 56 (1993).

\bibitem{jkt}
    M. Je\.zabek, J.H. K\"uhn and T. Teubner, \zpc{56}{1992}{653};
    M. Je\.zabek and T. Teubner, \zpc{59}{1993}{669}.

\bibitem{ms} H. Murayama and Y. Sumino, {\it Phys.~Rev.} {\bf D47}, 82
(1993).

\bibitem{r19}
P. Igo-Kemenes, M. Martinez, R. Miquel, and S. Orteu,
Talk given at {\it the Workshop on Physics and Experiments with
Linear $e^+e^-$ Colliders} (Waikoloa, Hawaii, April 1993).

\bibitem{r25}
    J.H. K\"uhn, in: F.A. Harris et al.~(eds.), 
    {\it Physics and Experiments with Linear $e^+e^-$ Colliders},
     Singapore: World Scientific, 1993, p.72.

\bibitem{r21}
K. Melnikov and O. Yakovlev,
{\it Phys. Lett.}~{\bf B324}, 217 (1994).

\bibitem{dthesis} 
    Y. Sumino, Ph.D.~Thesis, University of Tokyo preprint,
    UT-655 (1993).

\bibitem{r22} V. Fadin, V. Khoze, and A. Martin,
{\it Phys.~Rev.} {\bf D49}, 2247 (1994);
{\it Phys. Lett.}~{\bf B320}, 141 (1994).

\bibitem{khsj2}
V. Khoze and W. Sj\"{o}strand,
{\it Phys. Lett.}~{\bf B328}, 466 (1994).

\bibitem{fms} K. Fujii, T. Matsui and Y. Sumino,
{\it Phys.~Rev.} {\bf D50}, 4341 (1994).

\bibitem{r23}
W. M\"{o}dritsch and W. Kummer,
{\it Nucl. Phys.} {\bf B430}, 3 (1994);
W. Kummer and W. M\"{o}dritsch,
{\it Z. Phys.} {\bf C66}, 225 (1995).

\bibitem{cracow}
Y.~Sumino, {\it Acta Phys. Polonica}~{\bf B25} 1837 (1994).

\bibitem{r27}
    M. Je\.zabek, {\it Nucl.~Phys.~}{\bf 37 B} {\it(Proc.Suppl.)} (1994) 197.
\bibitem{r38}
    R. Harlander, M. Je\.zabek, J.H. K\"uhn and T. Teubner,
    \plb{346}{1995}{137}.
\bibitem{r52} 
    W. M\"odritsch, {\it Nucl. Phys.} {\bf B475}, 507 (1996).
\bibitem{cmmo}
    P.~Comas, R.~Miquel, M.~Martinez and S.~Orteu, in
edited by P.~Zerwas, 
{\it $e^+e^-$ Collisions at TeV Energies: The Physics Potential, Part D},
DESY 96-123D, (1996).

\bibitem{r26} R. Harlander, 
M. Je\.{z}abek, J. K\"{u}hn, and M.~Peter,
{\it Z.~Phys.} {\bf C73}, 477 (1997).

\bibitem{ps}
    M.~Peter and Y.~Sumino, hep-ph/9708223.

\bibitem{cdf}
    CDF Collaboration, Fermilab preprint, FERMILAB-PUB-97-284-E (1997). 
\bibitem{dzero}
    D0 Collaboration, {\it Phys.~Rev.~Lett.} {\bf 79}, 1197 (1997).

\bibitem{jk1}
    M. Je\.zabek and J.H. K\"uhn, \prd{48}{1993}{R1910}; erratum
    \prd{49}{1994}{4970}; and references therein.

\bibitem{kuehn}
    J.H. K\"uhn, \app{12}{1981}{347}.

\bibitem{kz} J.H. K\"uhn and P.M. Zerwas,
{\it Phys.~Rep.} {\bf 167}, 321 (1988).

\bibitem{hkmn}
K. Hagiwara, K. Kato, A.~D. Martin and C.-K. Ng, 
{\it Nucl.~Phys.}~{\bf B344}, 1 (1990).

\bibitem{hoang}
A.~Hoang, hep-ph/9704325.
\bibitem{hl}
A.~Hoang, P.~Labelle and S.~Zebarjad, hep-ph/9707337.

\bibitem{cl}
W.~Caswell and G.~Lepage, {\it Phys.~Rev.} {\bf A20}, 36 (1979).

\bibitem{labelle}
P.~Labelle, hep-ph/9608491; P~Labelle and S.~Zebarjad, hep-ph/9611313.
\bibitem{gr}
B.~Grinstein and I.~Rothstein, hep-ph/9703298.

\bibitem{r44}
    A. Czarnecki, M. Je\.zabek and J.H. K\"uhn,
    \npb{351}{1991}{70};\\
    A. Czarnecki and  M. Je\.zabek, \npb{427}{1994}{3}.
\bibitem{r28}
    M. Je\.zabek, \app{26}{1995}{789}.

\bibitem{ks}
J.H. K\"uhn and K.H. Streng, \npb{198}{1982}{71}.
\bibitem{r43}
    M. Je\.zabek and J.H. K\"uhn, \npb{320}{1989}{20}.

\bibitem{grzadkowski}
    See, for example, the talk given by B.~Grzadkowski in these 
    proceedings.

\end{thebibliography}
\end{document}